\documentclass{aa}  

\usepackage{graphicx}
\usepackage{txfonts}
\begin{document}

   \title{Bayesian evidence for two slow-wave damping models in hot coronal loops}

     \author{I. Arregui
          \inst{1,2}
          \and
          D. Y. Kolotkov\inst{3,4}
          \and
          V. M. Nakariakov\inst{3,5}
          }

   \institute{Instituto de Astrof\'{\i}sica de Canarias, E-38205 La Laguna, Tenerife, Spain\\
              \email{iarregui@iac.es}
         \and
            Departamento de Astrof\'{i}sica, Universidad de La Laguna, E-38206 La Laguna, Tenerife, Spain
          \and
          Centre for Fusion, Space and Astrophysics, Physics Department, University of Warwick, Coventry CV4 7AL, United Kingdom
          \and
          Engineering Research Institute \lq\lq Ventspils International Radio Astronomy Centre (VIRAC)\rq\rq\ of Ventspils University of Applied Sciences, Inzenieru iela 101, Ventspils, LV-3601, Latvia
          \and
            Centro de Investigacion en Astronom\'{\i}a, Universidad Bernardo O'Higgins, Avenida Viel 1497, Santiago, Chile\\
             }

   \date{Received ; accepted}

 
  \abstract{
We compute the evidence in favour of two models, one based on field-aligned thermal conduction alone and another that includes thermal misbalance as well, in explaining the damping of slow magneto-acoustic waves in hot coronal loops.  Our analysis is based on the computation of the marginal likelihood and the Bayes factor for the two damping models.  We quantify their merit in explaining the apparent relationship between slow mode periods and damping times, measured with SOHO/SUMER in a set of hot coronal loops. The results indicate evidence in favour of the model with thermal misbalance in the majority of the sample, with a small population of loops for which thermal conduction alone is more plausible.  The apparent possibility of two different regimes of slow-wave damping, if due to differences between the loops of host active regions and/or the photospheric dynamics, may help with revealing the coronal heating mechanism.}
  

   \maketitle
%

\section{Introduction}

Standing and propagating slow magnetohydrodynamic (MHD) waves in the solar corona have been extensively studied over the past two decades \citep[see e.g.][for comprehensive reviews]{demoortel05,roberts06,wang21}. A phenomenon that has attracted particular attention is the appearance of strongly damped Doppler-shift oscillations of ultraviolet emission lines in hot coronal loops ($>$ 6 MK), first reported by \cite{kliem02} and \cite{wang02a}, in observations with the SUMER spectrometer onboard SOHO.  The oscillations show up in hot lines, e.g. \ion{Fe}{XIX} and \ion{Fe}{XXI}, and are related to the hot plasma component of active region loops. Hot loops are typically observed in the X-ray band and in hot ultraviolet and extreme-ultraviolet lines \citep{reale14}. They correspond to those already identified in early rocket missions \citep{vaiana73}. Oscillations of similar nature were also detected with the Bragg Crystal Spectrometer on Yohkoh by \cite{mariska05,mariska06}. A quarter-period phase shift between intensity and Doppler-shift perturbations allowed for the interpretation of these observations as standing slow-mode magneto-acoustic waves.  The oscillations are frequently associated with small (or micro-) flares that have an occurrence rate of 3 to 14 per hour, and lifetimes that range from 5 to 150 min \citep{wang06}. Many events belong to recurring episodes, with a rate of 2-3 times within a couple of hours \citep{wang07,2011SSRv..158..397W}. The increase in the number of detected events has enabled to characterise their oscillatory properties statistically, finding periods and damping times in the ranges [10, 30] and [5, 35] min, respectively \citep{wang03b,wang03a}. These oscillations have a proven seismological potential, already demonstrated in applications to the inference of the magnetic field strength \citep{wang07} and the properties of the coronal plasma heating/cooling function \citep{kolotkov20}, for example.

Frequently invoked mechanisms to explain the observed rapid damping of coronal slow-mode waves include thermal conduction \citep{ofman02c, demoortel03}, compressive viscosity \citep{mendoza04,sigalotti07}, optically thin radiation \citep{pandey06}, nonlinear effects \citep{nakariakov00b} and their multiple combinations. Different mechanisms seem to be favoured depending on the damping regime (weak/strong), temperature, and density ranges.  A comprehensive overview of different physical scenarios for the damping of the fundamental mode of slow magneto-acoustic oscillations in coronal loops with different lengths, temperatures, and densities, under different mechanisms can be found in \cite{prasad21}. A table with a summary of proposed damping mechanisms and a discussion, based on the analysis of the scaling of the damping time with the wave period, is presented by \cite{wang21}. On the other hand, an almost linear scaling between the slow-wave damping times and oscillation periods, confidently observed in the solar \citep[see e.g.][]{2008ApJ...685.1286V, 2011SSRv..158..397W, nakariakov19} and stellar \citep{2016ApJ...830..110C} coronae up to periods of 30 min and even longer, can be explained with none of those damping mechanisms.

A mechanism that has recently gathered an increasing interest concerning the damping of slow MHD waves is the process of thermal misbalance, whereby compressive waves and a heated coronal plasma can exchange energy in a continuous interplay between wave-perturbed cooling and heating processes \citep{kolotkov19}. Thus, such a wave-induced thermal misbalance can enhance or suppress the damping of slow waves, depending on the parameter values of the heating/cooling model \citep{nakariakov17,kolotkov19,duckenfield21}. As shown by \cite{kolotkov19}, in the regime of enhanced damping, the theoretically obtained damping rates are about those estimated from SUMER observations of hot coronal loops. Furthermore, \cite{kolotkov22} and \citet{2021SoPh..296...20P} considered a model with field-aligned thermal conduction and wave-induced thermal misbalance to address the scaling of the damping time with period of standing slow waves in coronal loops observed in Doppler-shift with SUMER. In particular, \cite{kolotkov22} showed that accounting for the effect of thermal misbalance makes the relationship between the slow-wave damping time and period of a non-power-law form, unlike the damping mechanisms described above. 

In this paper, we quantify the evidence in favour of each of the two damping models considered by \cite{kolotkov22}: one with thermal conduction alone and the other with the addition of thermal misbalance. We compare the plausibility of the newly proposed thermal misbalance mechanism in front of thermal conduction, which is used as a reference model. The aim is to assess which mechanism explains better, completely or in part, the damping properties of slow magneto-acoustic waves in hot coronal loops in SUMER observations.

\section{Damping models}

The theoretical prediction for the damping time of slow magnetoacoustic waves due to field-aligned thermal conduction in a weakly dissipative limit can be expressed as \citep[e.g.][]{2014ApJ...789..118K, 2016ApJ...820...13M}
\begin{equation}\label{eq:tc}
\tau^{\rm {TC}}_{\rm D} =  \frac{2}{d} P^2, \, \mbox{with} \,\, d=\frac{4\pi^2(\gamma-1)k_{\|}}{\gamma \rho_0 C_{\rm v} c^2_{s}}.
\end{equation}
Here, $P$ is the wave period and $d$ is the thermal conduction parameter \citep{demoortel03}. We can fix the following set of physical parameters appearing in Eq.~(\ref{eq:tc}) using standard coronal values: the adiabatic index $\gamma = 5/3$, the field-aligned thermal conduction coefficient $k_\|  = 10^{-11} T_0^{5/2}$ [W\,m$^{-1}$\,K$^{-1}$] (with $T_0 = 6.3$\,MK a typical SUMER oscillation detection temperature, \citealt{wang02a}), the sound speed $c_{\rm s} = \sqrt{\gamma k_{\rm B} T_0 / m}$ (with $k_{\rm B}$ the Boltzmann constant and $m=0.6 \times 1.67\times 10^{-27}$ kg the mean particle mass), and the specific heat capacity $C_{\rm v}=(\gamma-1)k_{\rm B}/m$. This results in a model, $M_{\rm TC}$, with the plasma density $\rho_0$ as the only unknown, that we gather in the parameter vector $\mbox{\boldmath$\theta$}_{\rm TC} =\{\rho_0\}$.  For plasma densities in the range $\rho_0\in [0.5, 10]\times 10^{-12}$~kg\,m$^{-3}$, characteristic of hot coronal loops \citep{wang07}, values of $d$ in the range $d\sim[8,176]$ min are obtained, which leads to model predictions for the damping by thermal conduction in the range  
$\tau^{\rm TC}_{\rm D}\sim[1.1, 360] $~min, for periods $P$ between 10 and 40~min typically detected in observations.

An alternative theoretical prediction for the damping time of slow magnetoacoustic modes due to a combined effect of field-aligned thermal conduction and wave-induced thermal misbalance is 
\citep{kolotkov22}
\begin{equation}\label{eq:tm}
\tau^{\rm {TM}}_{\rm D} = \frac{2\tau_{\rm M}\, P^2}{d\tau_{\rm M} + P^2},
\end{equation}
with $\tau_{\rm M}$ being the thermal misbalance time determined by the properties of the coronal heating/cooling function. Equations~ (\ref{eq:tc}) and (\ref{eq:tm}) were derived under the assumption of weak dissipation, in which the ratios of oscillation period to thermal conduction and thermal misbalance times are small. In Eq.~(\ref{eq:tm}), $\tau^{\rm {TM}}_{\rm D}$ depends on two unknowns that we gather in the parameter vector $\mbox{\boldmath$\theta$}_{\rm TM} =\{\rho_0,\tau_\mathrm{M}\}$.  For this model, plasma densities in the range $\rho_0\in [0.5, 10]\times 10^{-12}$~kg\,m$^{-3}$ (as considered before) together with values of $\tau_{\rm M}$ in the range [1, 30]~min \citep[see e.g.][Fig. 2]{kolotkov20}, lead to damping times in the range $\tau^{\rm TM}_{\rm D}\sim[0.8, 47]$~min, for observed periods $P$ between 10 and 40~min.

\section{Evidence analysis and results}

Our analysis makes use of observations of standing slow waves in coronal loops observed in  Doppler-shift with SUMER. The whole SUMER spectral window contains a number of lines formed in the temperature range of 0.01--10 MK. They include the transition region line, \ion{S}{III}/\ion{Si}{III} at 
1113\AA\ (0.03--0.06 MK), the coronal lines \ion{Ca}{X} at 557\AA\ (0.7 MK) and \ion{Ca}{XIII}  at 1133\AA\ (2 MK), as well as the flare-lines \ion{Fe}{XIX} at 1118\AA\ (6.3 MK) and \ion{Fe}{xx}  at 567\AA\ (8 MK) \citep{wang02a}. We restrict our analysis to a selection of events corresponding to detections at 6.3 MK, from those summarised in \cite{2003A&A...406.1105W} and \cite{nakariakov19}. We deliberately use data obtained with the same instrument and observed at the same emission spectral line to exclude the temperature of the emitting plasma as a free parameter.
This selected SUMER observations were recently employed by \cite{kolotkov22} to validate Eq.~(\ref{eq:tm}) for the damping by thermal misbalance. In their analysis, \cite{kolotkov22} fix the plasma temperature to that of the SUMER observational channel in which most standing slow-wave events were observed (6.3\,MK). Treating the plasma density and the characteristic time-scale of thermal misbalance as free parameters, they find a reasonable agreement between theory and observations. However, a small number of data-points fall outside the region covered by the posterior predictive distribution of the samples obtained by Bayesian Markov Chain Monte Carlo (MCMC) sampling (see Figure 1 in \citealt{kolotkov22}), indicating that the effect of thermal misbalance is apparently less important in the slow-wave damping in those events.

To rigorously quantify the evidence of the two damping models given by Eqs.~(\ref{eq:tc}) and (\ref{eq:tm}) in explaining the set of observations, we follow a similar procedure to the one employed by \cite{arregui21} for the damping of transverse coronal loop oscillations,  based on the application of Bayesian model comparison \citep[see][for reviews on recent applications in the context of coronal seismology]{arregui22,anfinogentov22}. We first construct a two-dimensional grid over the synthetic data space $\mathcal{D}=(P, \tau_{\rm D})$, which covers the ranges in the oscillation period and damping time in observations.  The magnitude of the marginal likelihood for the two damping models over that space gives a measure of how well a particular period-damping time combination is predicted by each model. For the model with damping by thermal conduction $M_{\rm TC}$, with the parameter vector $\mbox{\boldmath$\theta$}_{\rm TC}$, the marginal likelihood is computed as
\begin{equation}\label{eq:ml}
p(\mathcal{D}|M_{\rm TC}) = \int {\rm d}\mbox{\boldmath$\theta$}_{\rm TC}\, p(\mathcal{D}|\mbox{\boldmath$\theta$}_{\rm TC},M_{\rm TC})\, p(\mbox{\boldmath$\theta$}_{\rm TC}|M_{\rm TC}),  
\end{equation}
and likewise for the model with damping by thermal misbalance, $M_{\rm TM}$, with the parameter vector  $\mbox{\boldmath$\theta$}_{\rm TM}$. 
The first factor in the integrand is the likelihood function. Under the assumption of a Gaussian likelihood function and adopting an error model for the damping time alone
\begin{equation}\label{eq:like}
p(\mathcal{D}|\mbox{\boldmath$\theta$}_{\rm TC}, M_{\rm TC}) =\frac{1} {\sqrt{2\pi} \sigma} 
\exp \Bigg\{-\frac{\left[\tau_{\rm D} - \tau^{\rm TC}_{\rm D}(\mbox{\boldmath$\theta$}_{\rm TC})\right]^2}{2\sigma^2}\Bigg\},
\end{equation}
and correspondingly for the thermal misbalance model. In this expression, $\sigma$ is the uncertainty in the damping time $\tau_{\rm D}$.  In the absence of specific values from the literature, this is fixed to the chosen value $\sigma= 0.1 \tau_{\rm D}$. Larger uncertainty values lead to lower levels of evidence. Possible data realisations from the two considered models are generated using the theoretical predictions given by Eqs.~(\ref{eq:tc}) and (\ref{eq:tm}) for models $M_{\rm TC}$ and $M_{\rm TM}$, respectively. 

The second factor in the integrand of Eq.~(\ref{eq:ml}) is the prior probability density of the model parameters. Based on the inference results obtained by \cite{kolotkov22}, we choose the following Gaussian priors for the two unknown parameters:   $\mathcal{G}(\rho_0[{10^{-12}\rm\,kg\, m}^{-3}], 4, 2)$ and $\mathcal{G}(\tau_{\rm M} [{\rm min}], 14.2, 5.0)$,  with the numerical values indicating the mean and the standard deviation, respectively.

\begin{figure*}[!t]
               \centering
         \includegraphics[scale=0.65]{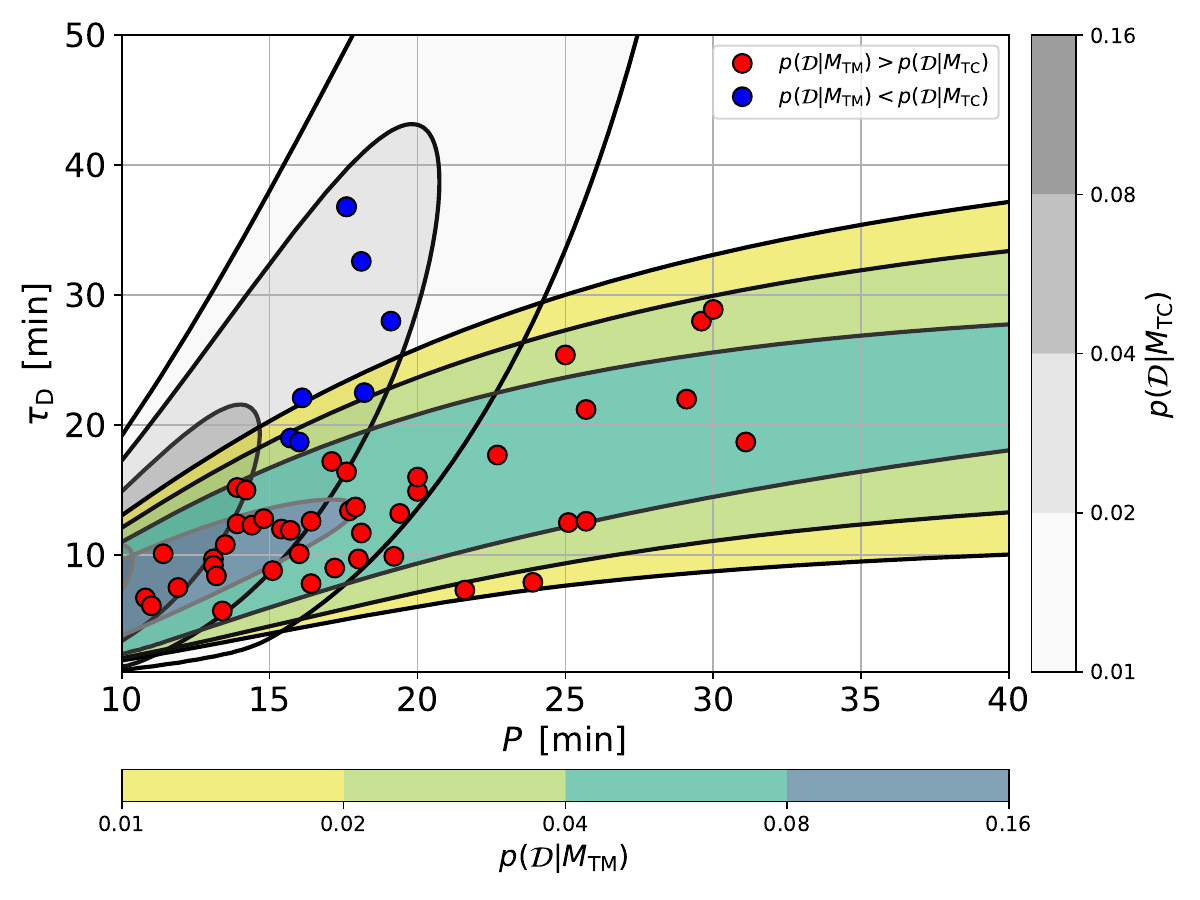}
          \includegraphics[scale=0.65]{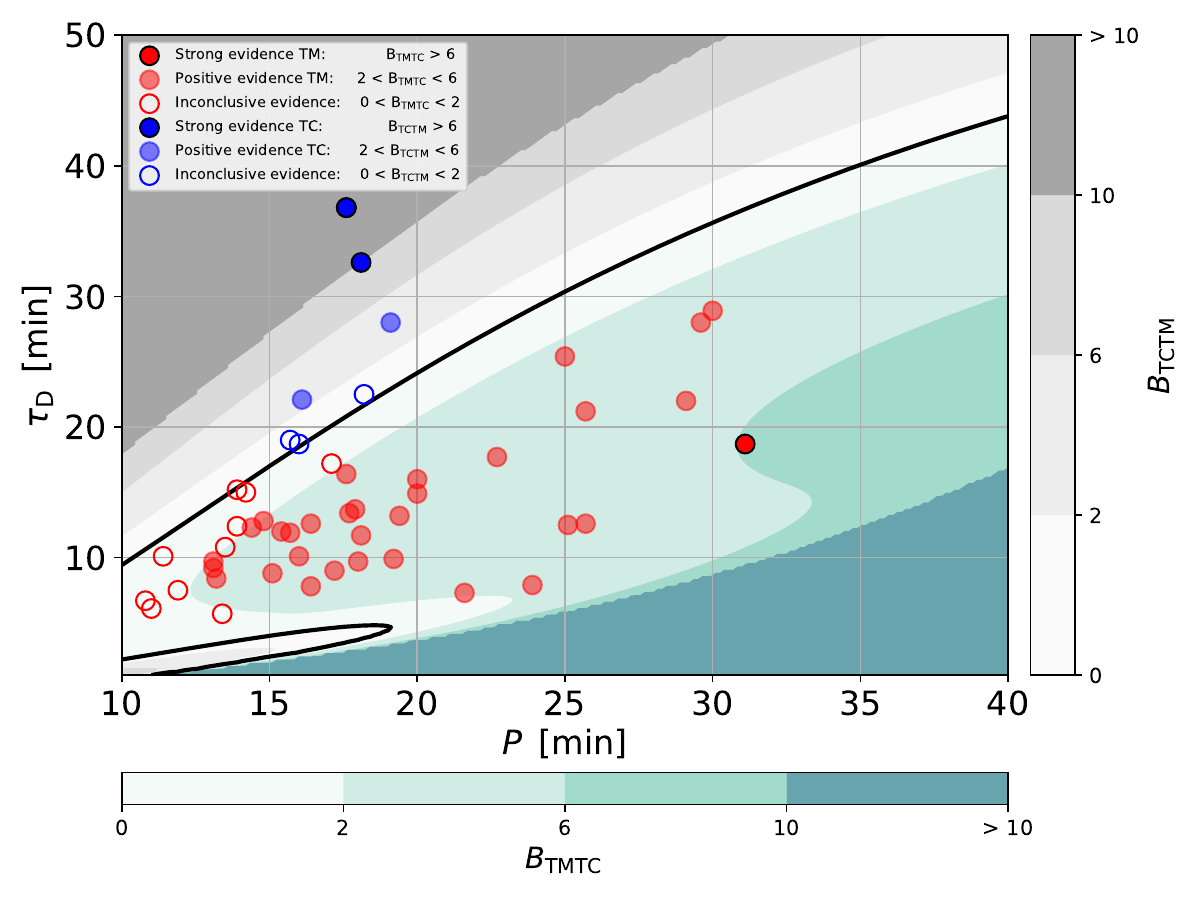}
\caption{Filled contour plot of the marginal likelihood values for the damping models $M_{\rm TC}$  and $M_{\rm TM}$ given by Eqs.~(\ref{eq:tc}) and (\ref{eq:tm}), respectively (top panel), and the corresponding Bayes factors (bottom panel) over the synthetic data space  $\mathcal{D}=(P, \tau_{\rm D})$. Equations~(\ref{eq:ml}) and (\ref{eq:bf}) are computed over a grid with $N_{P}=121$ and $N_{\tau_{\rm D}}=181$ points over the ranges $P\in[10,40]$ min and $\tau_{\rm D}\in[1,50]$ min. Over-plotted in circles are the observations corresponding to standing slow waves in coronal loops observed in  Doppler shift by SUMER at 6.3~MK. They are coloured according to the comparison between the marginal likelihood values (top panel) and the levels of evidence for each model (bottom panel) as indicated in the corresponding legends. \label{fig:f1}}
\end{figure*}

In Figure~\ref{fig:f1} (top), we show the resulting distribution of the marginal likelihood for the two compared models of the damping, over the grid of synthetic data in the damping time and oscillation period. Although the two marginal likelihood distributions do overlap, especially in the area with periods and damping times below 20 min, the contours associated to the different levels of evidence for each model and their shape can be clearly distinguished. Marginal likelihood contours for thermal conduction alone bend upwards towards regions with weaker damping  in the lower period range. Marginal likelihood contours for the model with thermal misbalance extend towards longer period values at comparatively lower damping time ranges. 
They nicely cover the grey-shaded area in Figure 1 by \cite{kolotkov22}, which represents the posterior predictive distribution obtained from the MCMC samples in their analysis. The figure also shows the location of the SUMER data plotted over the contours. Most of the data-points (42 out of 49) fall  over areas where the marginal likelihood for the model with thermal misbalance is larger than the marginal likelihood for the model with thermal conduction alone, and the evidence thus favours the former mechanism. For the remaining seven cases, the opposite happens and the marginal likelihood for the thermal conduction model is larger. 

To quantify the relative evidence between the two compared models, we assume that the two models are equally probable a priori, $p(M_{\rm TC}) = p(M_{\rm TM})$, and make use of the Bayes factor, given by
\begin{equation}\label{eq:bf}
B_{\rm TCTM} = 2\log \frac{p(\mathcal{D}|M_{\rm TC})}{p(\mathcal{D}|M_{\rm TM})} = -B_{\rm TMTC.}
\end{equation}
The assessment in terms of levels of evidence is based on the use of the empirical table by \cite{kass95} from the values thus obtained. The evidence in favour of model $M_{\rm TC}$ in front of model $M_{\rm TM}$ is deemed inconclusive for values of $B_{\rm TCTM}$ from 0 to 2; positive for values from 2 to 6; strong for values from 6 to 10; and very strong for values above 10. A similar tabulation applies to $B_{\rm TMTC}$.

Figure~\ref{fig:f1} (bottom) shows the corresponding Bayes factor distributions. By definition, the regions where $B_{\rm TCTM}$ and $B_{\rm TMTC}$ reach the different levels of evidence are mutually exclusive and cannot overlap.
They are clearly separated by the solid line that connects the points where $p(\mathcal{D}|M_{\rm TC}) = p(\mathcal{D}|M_{\rm TM})$ and thus the Bayes factors vanish, $B_{\rm TCTM} = B_{\rm TMTC}= 0$. In the surrounding white area, the Bayes factors are not large-enough to deem positive evidence in favour of any of the two damping models. Then, moving towards the top-left corner, the evidence favours thermal conduction with increasing levels of evidence. Moving towards the bottom-right corner, the evidence supports thermal misbalance with increasing levels of evidence. We can calculate numerical values for the Bayes factor for each of the 49 SUMER loop oscillation events. Differently coloured circles are used to represent different levels of evidence.  The majority of observed data points (32 out of 49) fall into the region where the evidence supports a model with wave-induced thermal misbalance in comparison to a model with thermal conduction alone. For several data-points (13 out of 49), the evidence is inconclusive (edge coloured circles).  In four cases, the evidence is positive (even strong in one of them) in favour of damping by thermal conduction alone. 
 
\section{Conclusion}

We considered two damping models for the explanation of the damping of standing slow magneto-acoustic waves in hot coronal loops, and computed the evidence in favour of each of them in explaining a set of observed oscillation periods and damping times in SUMER observations with measured periods and damping times.  We find a clear separation in the oscillation period and damping time data space between the regions with evidence in favour of each of the two models. The majority of the observed data points ($\sim$ 65\%) fall into the region where the evidence supports a model which links the oscillation damping with wave-induced thermal misbalance added to thermal conduction. Some data from the sample ($\sim$ 8\%) fall into the region where the evidence supports a damping model based on thermal conduction alone. These few cases may be regarded as a separate population of hot coronal loops for which particular physical or wave characteristics may make thermal conduction more plausible/dominant. 

The apparent possibility of two different regimes of slow oscillation damping could possibly be attributed to some variation of the coronal heating function in the events appearing in those different regimes. In the model which links the damping with thermal misbalance, we assumed that the radiative losses and the heating function are both uniquely determined by thermodynamic parameters of the plasma, i.e., the density and temperature. However, the yet unknown heating function could also depend on some other parameters which are not accounted by the model, for example, the energy supply flux which may vary in time. This would make the parameter $\tau_{\rm M}$ different in the events which belong to the two different populations (in particular, $\tau_{\rm M}\to \infty$ for the population better described by conductive damping alone). The identification of the differences between the loops and/or parameters of host active regions and/or the photospheric dynamics in the events which belong to different populations may shed light on the differences in the heating function, and help with revealing the coronal heating mechanism.  

\begin{acknowledgements}
This research was conducted while I.A. was a visitor at the Centre for Fusion, Space and Astrophysics, Department of Physics, University of Warwick. It is a pleasure for I.A. to acknowledge the financial support, the warm hospitality, and the friendly atmosphere during his visit.  I.A. is supported by project PID2021-127487NB-I00 from Ministerio de Ciencia, Innovaci\'on y Universidades and FEDER funds. D.Y.K. and V.M.N. acknowledge support from the STFC consolidated grant ST/T000252/1 and the Latvian Council of Science Project \lq\lq Multi-Wavelength Study of Quasi-Periodic Pulsations in Solar and Stellar Flares\rq\rq\ No. lzp-2022/1-0017.
SUMER is part of SOHO, the Solar and Heliospheric Observatory, of ESA and NASA. 
The SUMER project is financially supported by the Deutsches Zentrum f\"{u}r Luft- und Raumfahrt (DLR), the Centre National d'Etudes Spatiales (CNES), the National Aeronautics and Space Administration (NASA), and the European Space Agency's (ESA) PRODEX program (Swiss contribution).
\end{acknowledgements}


\begin{thebibliography}{37}
\expandafter\ifx\csname natexlab\endcsname\relax\def\natexlab#1{#1}\fi

\bibitem[{{Anfinogentov} {et~al.}(2022){Anfinogentov}, {Antolin}, {Inglis},
  {Kolotkov}, {Kupriyanova}, {McLaughlin}, {Nistic{\`o}}, {Pascoe}, {Krishna
  Prasad}, \& {Yuan}}]{anfinogentov22}
{Anfinogentov}, S.~A., {Antolin}, P., {Inglis}, A.~R., {et~al.} 2022, \ssr,
  218, 9

\bibitem[{Arregui(2021)}]{arregui21}
Arregui, I. 2021, The Astrophysical Journal Letters, 915, L25

\bibitem[{{Arregui}(2022)}]{arregui22}
{Arregui}, I. 2022, Frontiers in Astronomy and Space Sciences, 9, 826947

\bibitem[{{Cho} {et~al.}(2016){Cho}, {Cho}, {Nakariakov}, {Kim}, \&
  {Kumar}}]{2016ApJ...830..110C}
{Cho}, I.~H., {Cho}, K.~S., {Nakariakov}, V.~M., {Kim}, S., \& {Kumar}, P.
  2016, \apj, 830, 110

\bibitem[{{De Moortel}(2005)}]{demoortel05}
{De Moortel}, I. 2005, Royal Society of London Philosophical Transactions
  Series A, 363, 2743

\bibitem[{{De Moortel} \& {Hood}(2003)}]{demoortel03}
{De Moortel}, I. \& {Hood}, A.~W. 2003, \aap, 408, 755

\bibitem[{{Duckenfield} {et~al.}(2021){Duckenfield}, {Kolotkov}, \&
  {Nakariakov}}]{duckenfield21}
{Duckenfield}, T.~J., {Kolotkov}, D.~Y., \& {Nakariakov}, V.~M. 2021, \aap,
  646, A155

\bibitem[{{Kass} \& {Raftery}(1995)}]{kass95}
{Kass}, R.~E. \& {Raftery}, A.~E. 1995, JASA, 90, 773

\bibitem[{{Kliem} {et~al.}(2002){Kliem}, {Dammasch}, {Curdt}, \&
  {Wilhelm}}]{kliem02}
{Kliem}, B., {Dammasch}, I.~E., {Curdt}, W., \& {Wilhelm}, K. 2002, \apjl, 568,
  L61

\bibitem[{{Kolotkov} {et~al.}(2020){Kolotkov}, {Duckenfield}, \&
  {Nakariakov}}]{kolotkov20}
{Kolotkov}, D.~Y., {Duckenfield}, T.~J., \& {Nakariakov}, V.~M. 2020, \aap,
  644, A33

\bibitem[{{Kolotkov} \& {Nakariakov}(2022)}]{kolotkov22}
{Kolotkov}, D.~Y. \& {Nakariakov}, V.~M. 2022, \mnras, 514, L51

\bibitem[{{Kolotkov} {et~al.}(2019){Kolotkov}, {Nakariakov}, \&
  {Zavershinskii}}]{kolotkov19}
{Kolotkov}, D.~Y., {Nakariakov}, V.~M., \& {Zavershinskii}, D.~I. 2019, \aap,
  628, A133

\bibitem[{{Krishna Prasad} {et~al.}(2014){Krishna Prasad}, {Banerjee}, \& {Van
  Doorsselaere}}]{2014ApJ...789..118K}
{Krishna Prasad}, S., {Banerjee}, D., \& {Van Doorsselaere}, T. 2014, \apj,
  789, 118

\bibitem[{{Mandal} {et~al.}(2016){Mandal}, {Magyar}, {Yuan}, {Van
  Doorsselaere}, \& {Banerjee}}]{2016ApJ...820...13M}
{Mandal}, S., {Magyar}, N., {Yuan}, D., {Van Doorsselaere}, T., \& {Banerjee},
  D. 2016, \apj, 820, 13

\bibitem[{{Mariska}(2005)}]{mariska05}
{Mariska}, J.~T. 2005, \apjl, 620, L67

\bibitem[{{Mariska}(2006)}]{mariska06}
{Mariska}, J.~T. 2006, \apj, 639, 484

\bibitem[{{Mendoza-Brice{\~n}o} {et~al.}(2004){Mendoza-Brice{\~n}o},
  {Erd{\'e}lyi}, \& {Sigalotti}}]{mendoza04}
{Mendoza-Brice{\~n}o}, C.~A., {Erd{\'e}lyi}, R., \& {Sigalotti}, L. D.~G. 2004,
  \apj, 605, 493

\bibitem[{{Nakariakov} {et~al.}(2017){Nakariakov}, {Afanasyev}, {Kumar}, \&
  {Moon}}]{nakariakov17}
{Nakariakov}, V.~M., {Afanasyev}, A.~N., {Kumar}, S., \& {Moon}, Y.~J. 2017,
  \apj, 849, 62

\bibitem[{{Nakariakov} {et~al.}(2019){Nakariakov}, {Kosak}, {Kolotkov},
  {Anfinogentov}, {Kumar}, \& {Moon}}]{nakariakov19}
{Nakariakov}, V.~M., {Kosak}, M.~K., {Kolotkov}, D.~Y., {et~al.} 2019, \apjl,
  874, L1

\bibitem[{{Nakariakov} {et~al.}(2000){Nakariakov}, {Verwichte}, {Berghmans}, \&
  {Robbrecht}}]{nakariakov00b}
{Nakariakov}, V.~M., {Verwichte}, E., {Berghmans}, D., \& {Robbrecht}, E. 2000,
  \aap, 362, 1151

\bibitem[{{Ofman} \& {Wang}(2002)}]{ofman02c}
{Ofman}, L. \& {Wang}, T. 2002, \apjl, 580, L85

\bibitem[{{Pandey} \& {Dwivedi}(2006)}]{pandey06}
{Pandey}, V.~S. \& {Dwivedi}, B.~N. 2006, \solphys, 236, 127

\bibitem[{{Prasad} {et~al.}(2021{\natexlab{a}}){Prasad}, {Srivastava}, \&
  {Wang}}]{prasad21}
{Prasad}, A., {Srivastava}, A.~K., \& {Wang}, T. 2021{\natexlab{a}}, \solphys,
  296, 105

\bibitem[{{Prasad} {et~al.}(2021{\natexlab{b}}){Prasad}, {Srivastava}, \&
  {Wang}}]{2021SoPh..296...20P}
{Prasad}, A., {Srivastava}, A.~K., \& {Wang}, T.~J. 2021{\natexlab{b}},
  \solphys, 296, 20

\bibitem[{Reale(2014)}]{reale14}
Reale, F. 2014, Living Reviews in Solar Physics, 11, 4

\bibitem[{{Roberts}(2006)}]{roberts06}
{Roberts}, B. 2006, Philosophical Transactions of the Royal Society of London
  Series A, 364, 447

\bibitem[{{Sigalotti} {et~al.}(2007){Sigalotti}, {Mendoza-Brice{\~n}o}, \&
  {Luna-Cardozo}}]{sigalotti07}
{Sigalotti}, L. D.~G., {Mendoza-Brice{\~n}o}, C.~A., \& {Luna-Cardozo}, M.
  2007, \solphys, 246, 187

\bibitem[{{Vaiana} {et~al.}(1973){Vaiana}, {Krieger}, \& {Timothy}}]{vaiana73}
{Vaiana}, G.~S., {Krieger}, A.~S., \& {Timothy}, A.~F. 1973, \solphys, 32, 81

\bibitem[{{Verwichte} {et~al.}(2008){Verwichte}, {Haynes}, {Arber}, \&
  {Brady}}]{2008ApJ...685.1286V}
{Verwichte}, E., {Haynes}, M., {Arber}, T.~D., \& {Brady}, C.~S. 2008, \apj,
  685, 1286

\bibitem[{{Wang}(2011)}]{2011SSRv..158..397W}
{Wang}, T. 2011, \ssr, 158, 397

\bibitem[{{Wang} {et~al.}(2021){Wang}, {Ofman}, {Yuan}, {Reale}, {Kolotkov}, \&
  {Srivastava}}]{wang21}
{Wang}, T., {Ofman}, L., {Yuan}, D., {et~al.} 2021, \ssr, 217, 34

\bibitem[{{Wang} {et~al.}(2007){Wang}, {Innes}, \& {Qiu}}]{wang07}
{Wang}, T.~J., {Innes}, D.~E., \& {Qiu}, J. 2007, \apj, 656, 598

\bibitem[{{Wang} {et~al.}(2006){Wang}, {Innes}, \& {Solanki}}]{wang06}
{Wang}, T.~J., {Innes}, D.~E., \& {Solanki}, S.~K. 2006, \aap, 455, 1105

\bibitem[{{Wang} {et~al.}(2002){Wang}, {Solanki}, {Curdt}, {Innes}, \&
  {Dammasch}}]{wang02a}
{Wang}, T.~J., {Solanki}, S.~K., {Curdt}, W., {Innes}, D.~E., \& {Dammasch},
  I.~E. 2002, \apjl, 574, L101

\bibitem[{{Wang} {et~al.}(2003{\natexlab{a}}){Wang}, {Solanki}, {Curdt},
  {Innes}, {Dammasch}, \& {Kliem}}]{wang03a}
{Wang}, T.~J., {Solanki}, S.~K., {Curdt}, W., {et~al.} 2003{\natexlab{a}},
  \aap, 406, 1105

\bibitem[{{Wang} {et~al.}(2003{\natexlab{b}}){Wang}, {Solanki}, {Curdt},
  {Innes}, {Dammasch}, \& {Kliem}}]{2003A&A...406.1105W}
{Wang}, T.~J., {Solanki}, S.~K., {Curdt}, W., {et~al.} 2003{\natexlab{b}},
  \aap, 406, 1105

\bibitem[{{Wang} {et~al.}(2003{\natexlab{c}}){Wang}, {Solanki}, {Innes},
  {Curdt}, \& {Marsch}}]{wang03b}
{Wang}, T.~J., {Solanki}, S.~K., {Innes}, D.~E., {Curdt}, W., \& {Marsch}, E.
  2003{\natexlab{c}}, \aap, 402, L17

\end{thebibliography}

\end{document}